\documentclass[a4paper,11pt]{article}
\usepackage{pos}

\usepackage{amsmath}
\usepackage{graphicx}

\newcommand{\bbR}{{\mathbb R}}
\newcommand{\bbC}{{\mathbb C}}
\newcommand{\comment}[1]{}

\title{Quantum tunneling in the real-time path integral by the Lefschetz
thimble method}

\author*[a,b]{Jun Nishimura}

\affiliation[a]{KEK Theory Center, Institute of Particle and Nuclear Studies,\\
High Energy Accelerator Research Organization,\\
1-1 Oho, Tsukuba, Ibaraki 305-0801, Japan}

\affiliation[b]{Graduate Institute for Advanced Studies, SOKENDAI,\\
1-1 Oho, Tsukuba, Ibaraki 305-0801, Japan}

\emailAdd{jnishi@post.kek.jp}

\abstract{Quantum tunneling is mostly discussed in the Euclidean path integral formalism using instantons. On the other hand, it is difficult to understand quantum tunneling based on the real-time path integral due to its oscillatory nature, which causes the notorious sign problem. We show that recent development of the Lefschetz thimble method enables us to investigate this issue numerically. In particular, we find that quantum tunneling occurs due to complex trajectories, which are actually observable experimentally by using the so-called weak measurement.}

\FullConference{%
  Corfu Summer Institute 2022 "School and Workshops on Elementary Particle Physics and Gravity",\\
  28 August - 1 October, 2022\\
  Corfu, Greece}

\begin{document}
\maketitle


\section{Introduction}

Quantum tunneling has been mostly discussed
in the imaginary-time path integral formalism
using instantons \cite{coleman_1985,ColemanPaperI,ColemanPaperII},
which enables us to 
investigate interesting phenomena such as
the decay rate of a false vacuum in quantum field theory (QFT),
bubble nucleation in first order phase transitions and domain wall fusions.
The tunneling amplitude one obtains in this way is suppressed in general
by $e^{-S_0/\hbar}$ with $S_0$ being the Euclidean action
for the instanton configuration,
which reveals its genuinely nonperturbative nature.

A natural question to ask here is
how to describe quantum tunneling directly in the real-time path integral.
This is motivated since, in reality, there are also contributions from 
classical motion over the barrier (c.f., sphalerons in QFT), which cannot
be taken into account by instantons.
Also the real-time path integral is needed to
obtain the quantum state after tunneling 
and its subsequent time-evolution.
However, a naive analytic  continuation of instantons 
leads to singular complex trajectories \cite{Cherman:2014sba}.
We clarify this issue completely by explicit Monte Carlo calculations.

The main obstacle in performing first-principle calculations
in the real-time path integral by using a Monte Carlo method
is
the severe sign problem, which occurs due to the
integrand involving
an oscillating factor $e^{iS[x(t)]}$, where the action $S[x(t)] \in \bbR$
depends on the path $x(t)$.
In order to overcome this problem,
we use the generalized Lefschetz thimble method (GTM) \cite{Alexandru:2015sua},
which was developed along the earlier proposals \cite{Fujii:2013sra,Cristoforetti:2012su,Witten:2010cx}.

Recently there have been further important developments of this method.
First, an efficient algorithm to generate a new configuration
was developed
based on
the Hybrid Monte Carlo algorithm (HMC), which is
applied to the variables
after the flow \cite{Fujii:2013sra,Fukuma:2019uot}
or before the flow \cite{Fujisawa:2021hxh}.
The former has an advantage that the modulus of the Jacobian
associated with the change of variables is included in the HMC procedure
of generating
a new configuration, whereas the latter has an advantage that
the HMC procedure
simplifies drastically
without increasing the cost
as far as one uses the backpropagation to calculate
the HMC force.
Second, the integration
of the flow time within an appropriate range
has been proposed \cite{Fukuma:2020fez}
to overcome the multi-modality problem that occurs when there are contributions
from multiple thimbles that are far separated from each other in the
configuration space.
Third, it has been realized that,
when the system size becomes large,
there is a problem that occurs
in solving the anti-holomorphic gradient flow equation,
which can be cured by optimizing the
flow equation with a kernel acting on the
drift term \cite{precondition2022}.

Here we apply the GTM
to the real-time path integral for the transition amplitude
in a simple quantum mechanical system,
where the use of various new techniques mentioned above turns out to
be crucial \cite{Nishimura:2023dky}.
This, in particular,
enables us to identify the relevant complex saddle points that
contribute to the path integral from first principles, which was not
possible before.
%
By introducing a large enough momentum in the initial wave function,
we find that the saddle point
becomes close to real,
which clearly indicates
the transition to classical dynamics.

In fact, the complex trajectories can be probed \cite{TANAKA2002307,Turok:2013dfa} by
the ensemble average of the coordinate $x(t)$ at time $t$,
which gives the ``weak value'' \cite{Aharonov:1988xu}
of the Hermitian coordinate operator $\hat{x}$
evaluated at time $t$ with a post-selected final wave
function.
Note that this is actually a physical quantity
that can be measured by experiments (``weak measurement'')
at least in principle. 
We calculate this quantity by taking the ensemble average numerically
and reproduce the result obtained by solving the Schr\"odinger equation,
which confirms the validity of our calculations.

The rest of this article is organized as follows.
In section \ref{section:review} we briefly review some previous works
in the case of a double-well potential, which will be important in our analysis.
In section \ref{section:method}
we explain the details of the calculation method
used in applying the GTM to the real-time path integral.
In section \ref{section:complex_traj_from_montecarlo}
we show our main result thus obtained.
In particular,
we identify relevant complex saddle points,
which are responsible for quantum tunneling.
Section \ref{section:summary} is devoted to a summary and discussions.


\section{Brief review of previous works}
\label{section:review}

In this section
we review some previous works on
a quantum system with a double-well potential
\begin{equation}
    V(x)=\lambda (x^2-a^2)^2 \ ,
    \label{eq:double-well_potential}
\end{equation}
which is a typical example used to discuss quantum tunneling.
Here we take $\lambda=1/2$ and $a=1$ 
in the potential \eqref{eq:double-well_potential}.
The Lagrangian is given by 
\begin{alignat}{3}
  L =
  \left( \frac{d x}{d t}\right)^2  - (x^2 -1 ) ^2  \ ,
  \label{eq:Lagrangian}
\end{alignat}
up to an overall factor, where the mass is set to unity.

First we review Ref.~\cite{Cherman:2014sba},
which discusses
the analytic continuation of the instanton solution
in the imaginary-time formalism.
Then we review Ref.~\cite{Tanizaki:2014xba},
in which all the solutions to the classical equation of motion
were obtained analytically
although it was not possible to identify
the relevant complex solutions
from the viewpoint of the Picard-Lefschetz theory.


\subsection{Analytic continuation of the instanton}
\label{sec:anal-cont-instanton}

Here we discuss a complex classical solution that can be obtained by
analytic continuation of the instanton solution
in the imaginary-time formalism \cite{Cherman:2014sba}.

\begin{figure}
    \centering
    \includegraphics[scale=0.6]{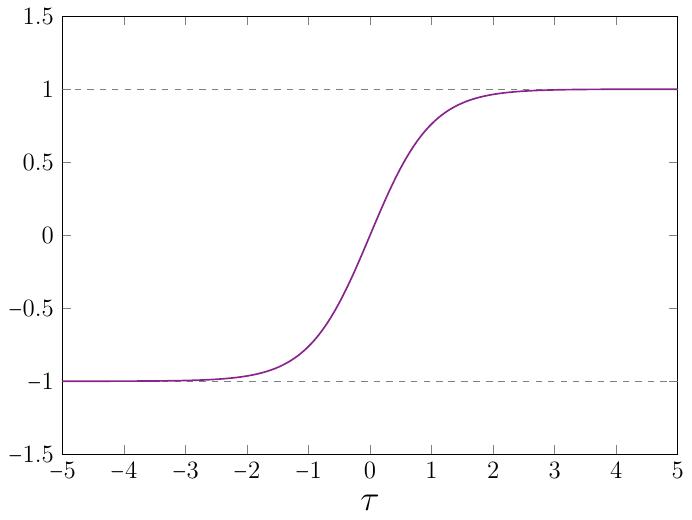}
    \includegraphics[scale=0.6]{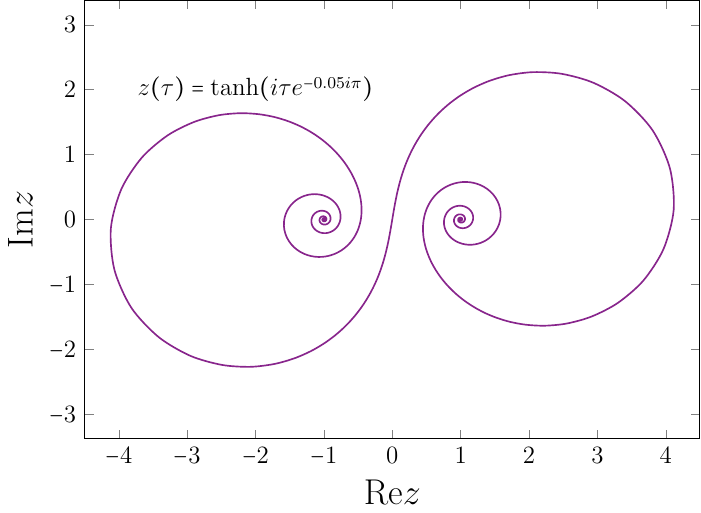}
    \caption{(Left) The instanton solution \eqref{eq:imaginary_instanton}
      in the imaginary time formalism ($\alpha=\frac{\pi}{2}$), which
      connects the two potential minima indicated by
      the horizontal dashed line.
      (Right) The trajectory of the
      complex solution
      obtained by analytic continuation from the instanton solution is
      shown in the complex plane for $\alpha=0.1 \times \frac{\pi}{2}$.}
    \label{fig:imaginary_instanton}
\end{figure}

For that, we consider the Wick rotation $t= e^{-i\alpha} \tau$,
where $\tau \in \bbR$ runs from $-\infty$ to $\infty$.
In particular, $\alpha=0$ corresponds
to the real time and $\alpha=\pi/2$ corresponds the imaginary time.
The Lagrangian \eqref{eq:Lagrangian} becomes
\begin{alignat}{3}
  L^{(\alpha)} =
 e^{2 i\alpha} \left( \frac{d z}{d \tau }\right)^2  - (z^2 -1 ) ^2  \ ,
\end{alignat}
where $z(t)$ represents a complex path.
For $\alpha=\frac{\pi}{2}$,
we obtain a real solution
\begin{equation}
  z^\star(\tau)=\tanh
  \tau \ ,
    \label{eq:imaginary_instanton}
\end{equation}
which satisfies the boundary condition
\begin{align}
  z^\star(-\infty)=-1 \ , \quad z^\star(\infty)=1
  \label{instanton-bc}
\end{align}
and therefore connects the two potential minima
as we plot in Fig.~\ref{fig:imaginary_instanton} (Left).
This is the instanton solution in the imaginary-time formalism,
and it actually describes quantum tunneling as is discussed, for instance,
in Ref.~\cite{coleman_1985}.

By making an analytic continuation from \eqref{eq:imaginary_instanton},
we can obtain a solution
for arbitrary $0<\alpha \le \frac{\pi}{2}$,
which is given by 
\begin{align}
    z^\star(\tau) = \tanh \Big( i \tau  e^{-i\alpha} \Big)
  \label{instanton-anal-continued}
\end{align}
satisfying the same boundary condition \eqref{instanton-bc}.
Note that this solution is complex for $\alpha<\frac{\pi}{2}$
and it gives a trajectory with a spiral shape as shown in
Fig.~\ref{fig:imaginary_instanton} (Right) 
for $\alpha=0.1 \times \frac{\pi}{2}$.
For smaller and smaller $\alpha$, the trajectory winds 
more and more 
around the potential minima
as $\tau \rightarrow \pm \infty$ and it also extends
farther and farther in the complex plane.
Thus the solution that can be obtained by analytic continuation from
the instanton is actually singular in the $\alpha \rightarrow 0$ limit.

On the other hand,
if one plugs \eqref{instanton-anal-continued}
in the action $S= \int dt L = e^{-i \alpha} \int d \tau L^{(\alpha)}$,
one finds that the $\tau$ integration for different $\alpha$
is related to each other by just rotating the integration contour of $\tau$
in the complex plane,
which implies that
the action
is independent of $\alpha$
due to Cauchy's theorem.
Therefore, the amplitude one obtains for this solution
in the $\alpha\rightarrow 0$ limit is suppressed by
$e^{iS[z^\star]/\hbar}=e^{-S_{\rm E}/\hbar}$,
where $S_{\rm E}>0$ is the Euclidean action for the instanton solution
\eqref{eq:imaginary_instanton},
which implies that the amplitude can be correctly reproduced
by the complex saddle point obtained in this way
as far as one introduces an infinitesimal $\alpha>0$ as a kind of regulator.

\subsection{Exact classical solutions for a double-well potential}
\label{sec:exact-solutions}

From the Lagrangian \eqref{eq:Lagrangian},
one can derive
the complex version of the energy conservation
\begin{equation}
    \left(\frac{dz}{dt}\right)^2+(z^2-1)^2=q^2 \ ,
    \label{eq:double-well_energy_conservation}
\end{equation}
where $q$ is some complex constant. 
This differential equation can be readily solved as
%
\begin{equation}
  z(t)=\sqrt{\frac{q^2-1}{2q}} \,
  \text{sd}
\left(\sqrt{2q}(t+c),\sqrt{\frac{1+q}{2q}}\right) \ ,
\label{zt-sol}
\end{equation}
where $c$ is another complex constant
and $\text{sd}(x,k)$ is the Jacobi elliptic function.
Thus the general classical solution
can be parametrized by
the two integration constants $q$ and $c$.

Let us fix the end points of the solution to be $z(0)=x_\text{i}$ and 
$z(T)=x_\text{f}$, which can be complex in general.
Then one finds that the parameter $k=\sqrt{(1+q)/2q}$,
which is called the elliptic modulus,
must satisfy the condition \cite{Tanizaki:2014xba}
\begin{align}
&    n\omega_1(k) +m\omega_3(k) \nonumber\\
& =\frac{T}{2}+\sqrt{\frac{2k^2-1}{8}}
   \left[\text{sd}^{-1}
   \left(\frac{\sqrt{2k^2-1}}{k\sqrt{2-2k^2}}x_\text{i},k\right)
 -(-)^{n+m}\text{sd}^{-1}
   \left(\frac{\sqrt{2k^2-1}}{k\sqrt{2-2k^2}}x_\text{f},k\right)\right]
    \label{eq:KT_constraint}
\end{align}
with $\omega_1(k)$ and $\omega_3(k)$ defined by
\begin{align}
    \omega_1(k)&=K(k)\sqrt{k^2-\frac{1}{2}} \ ,\\
    \omega_3(k)&=iK\left(\sqrt{1-k^2}\right)\sqrt{k^2-\frac{1}{2}} \ ,
\end{align}
where $K(k)$ is the complete elliptic integral of the first kind.
Note that the solution after fixing the end points still depends on 
two integers $(n,m)$, which we will refer to as modes of the solution
in what follows. 
Once we have $k$ and hence $q=1/(2k^2-1)$ for a given mode $(m,n)$, 
we can determine the complex parameter $c$ in \eqref{zt-sol}
from $z(0)=x_\text{i}$.

For each solution $z(t)$ obtained above,
we can obtain a solution $ \tilde{z}(t)= a z(a\sqrt{2\lambda}t)$
for arbitrary $\lambda$ and $a$ in \eqref{eq:double-well_potential}
that satisfies the boundary conditions $\tilde{z}(0)=a x_\text{i}$ 
and $\tilde{z}(T)=a x_\text{f}$.


\section{The calculation method used in this work}
\label{section:method}

In this section we explain how to perform Monte Carlo calculations
for the real-time path integral.
First we briefly review the basic idea of the GTM
to solve the sign problem.
Then we review the idea of integrating the flow time
to solve the multi-modality problem.
Finally we explain the problem of the anti-holomorphic gradient flow
that occurs in large systems, and discuss how to solve 
it by optimizing the gradient flow.

\subsection{The basic idea of the GTM}
\label{section:GTM}

In this section we give a brief review of the GTM,
which is a promising method for solving the sign problem
based on 
the Picard-Lefschetz theory.
Here we consider a general model defined by the partition function
and the observable
\begin{equation}
  Z=\int  d^Nx \, e^{-S(x)} \ , \quad
  \langle \mathcal{O} \rangle
  =\frac{1}{Z} \int  d^Nx \, \mathcal{O}(x) \, e^{-S(x)} \ , 
    \label{eq:thible_integral_appendix}
\end{equation}
where $x = (x_1 , x_1 , \cdots , x_N) \in \bbR^N$
and $d^N x = \prod_{n=1}^{N} dx_{n}$.
The action $S(x)$ is a complex-valued holomorphic function of $x$,
which makes \eqref{eq:thible_integral_appendix}
a highly oscillating multi-dimensional integral
and hence causes the sign problem when the number $N$ of variables
becomes large.
%
%

Let us first recall that the Picard-Lefschetz theory
makes the oscillating integral well-defined
by deforming the integration
contour
using the anti-holomorphic gradient flow equation
\begin{align}
    \frac{d z_i(\sigma)}{d\sigma}
=\overline{\frac{\partial S(z(\sigma))}{\partial z_i}}
\label{anti-hol-grad-flow-general}
\end{align}
with the initial condition $z(0)=x\in\mathbb{R}^N$,
where $\sigma$ plays the role of the deformation parameter.
This flow equation defines a one-to-one map
from $x=z(0) \in \bbR^N$ to $z=z(\tau)\in \mathcal{M}_\tau \in \bbC^N$.
Due to Cauchy's theorem, 
the partition function and the observable
\eqref{eq:thible_integral_appendix}
can be rewritten as
\begin{equation}
  Z
  = \int_{\mathcal{M}_\tau}d^Nz \, e^{-S(z)} \ , \quad
  \langle \mathcal{O} \rangle
  =\frac{1}{Z}\int_{\mathcal{M}_\tau}  d^Nz \, \mathcal{O}(z) \, e^{-S(x)} \ . 
    \label{eq:thible_integral_appendix2}
\end{equation}

The important property of the anti-holomorphic gradient flow 
equation \eqref{anti-hol-grad-flow-general} is that
\begin{equation}
    \frac{d S(z(\sigma))}{d\sigma}
=\sum_i \frac{\partial S(z(\sigma))}{\partial z_i}
\frac{d z_i(\sigma)}{d \sigma}
=\sum_i \left|\frac{\partial S(z(\sigma))}{\partial z_i}\right|^2
\geq 0 \ ,
\label{prop-anti-hol-flow}
\end{equation}
which means that 
the imaginary part of the effective action
is constant along the flow,
whereas the real part 
$S(z(\sigma))$ keeps on growing with $\sigma$
unless one reaches some saddle point $z=z^\star$ defined by
\begin{align}
\frac{\partial S(z^\star)}{\partial z_i} &= 0 \ .
\label{saddle-point-equation}
\end{align}
Thus, in the $\tau\rightarrow\infty$ limit, 
the manifold $\mathcal{M}_\tau$ is decomposed into 
the so-called Lefschetz thimbles, each of which is associated with
some saddle point.
The saddle points one obtains in this way are called 
``relevant''
in the Picard-Lefschetz theory.
In particular, the saddle points on the original integration contour
are always relevant.
Note also that there can be many saddle points that are not obtained
by deforming the original contour in this way, which are called
``irrelevant''.



In the $\tau \rightarrow \infty$ limit,
${\rm Im} \, S(z)$ becomes constant on each Lefschetz thimble
due to the property \eqref{prop-anti-hol-flow}
so that the sign problem is solved except for
the one\footnote{The sign problem due to
  the complex integration measure $d^Nz$ is called
  the residual sign problem. The severeness of this problem
  depends on the model and its parameters \cite{Fujii:2013sra}.}
 coming from the measure $d^Nz$.
 In the GTM \cite{Alexandru:2015sua},
 the flow time $\tau \rightarrow \infty$ limit is not taken,
 which has a big advantage over the
 earlier proposals \cite{Fujii:2013sra,Cristoforetti:2012su,Witten:2010cx}
 with $\tau =\infty$,
 which require 
 prior knowledge of the relevant saddle points.

The sign problem can still be ameliorated by choosing $\tau \sim \log N$,
which makes 
the reweighting method work.
However, the large flow time $\tau$
causes
the multi-modality problem
(or the ergodicity problem)
since the transitions among different regions of $\mathcal{M}_\tau$
that flow into different thimbles in the $\tau \rightarrow \infty$ limit
are highly suppressed during the simulation.

\subsection{Integrating the flow time}
\label{section:integrating-flow-time}

In order to solve both the sign problem and the multi-modality problem,
it was proposed \cite{Fukuma:2020fez}
to integrate the flow time $\tau$ as
\begin{equation}
  Z_W
  = \int_{\tau_{\rm min}}^{\tau_{\rm max}} d\tau  \, e^{- W(\tau)}
  \int_{\mathcal{M}_\tau}d^Nz \, e^{-S(z)}
    \label{eq:thible_integral_appendix3}
\end{equation}
with some weight $W(\tau)$, which is chosen to 
make
the $\tau$-distribution
roughly uniform in the region $ [\tau_{\rm min} , \tau_{\rm max}]$.
The use of this idea is important in our work since we have to
be able to sample all the saddle points and the associated thimbles
that contribute to the path integral.


For an efficient sampling in \eqref{eq:thible_integral_appendix3},
we
use the Hybrid Monte Carlo algorithm \cite{Duane:1987de},
which updates the configuration by solving a fictitious classical
Hamilton dynamics treating ${\rm Re} S(z)$ as the potential.
When we apply this idea to \eqref{eq:thible_integral_appendix3},
there are actually two options.

One option is to
define a fictitious classical Hamilton dynamics for
$(z,\tau)\in \mathcal{R}$ with $z \in \mathcal{M}_\tau$,
where $\mathcal{R}$ is the ``worldvolume'' obtained
by the foliation of $\mathcal{M}_\tau$ with
$\tau \in [\tau_{\rm min} , \tau_{\rm max}]$.
While this option has an important
advantage (See footnote \ref{footnote:modulus}.),
one has to treat
a system constrained on the worldvolume,
which makes the algorithm quite complicated.
Another problem is that the worldvolume is pinched if
there is a saddle point on the original integration contour,
which causes the ergodicity problem.


Here we adopt the other option, which is to rewrite
\eqref{eq:thible_integral_appendix3} as
\begin{equation}
  Z_W
  = \int_{\tau_{\rm min}}^{\tau_{\rm max}} d\tau  \, e^{- W(\tau)}
   \int d^Nx \,
   \det J(x,\tau) \, e^{-S(z(x,\tau))} \ ,
   \label{Z-x-parameter-space}
\end{equation}
where 
$z(x,\tau)$ represents the configuration obtained after the flow
starting from $x \in \bbR^N$
and
\begin{align}
  J_{ij}(x , \tau) =  \frac{\partial z_i(x , \tau) }{ \partial x_j}
\label{eq:def-Jacobi-matrix}
\end{align}
is the Jacobi matrix associated with the change of variables.
Then one can define a fictitious classical Hamilton dynamics for
$(x,\tau)\in \bbR^N \times [\tau_{\rm min} , \tau_{\rm max}]$.
Here
one only has to deal with an unconstrained
system, which makes the algorithm simple.
The disadvantage, however, is that the Jacobian $\det J(x,\tau)$
that appears in \eqref{Z-x-parameter-space}
has to be taken into
account by reweighting, which causes the
overlap problem\footnote{Note that
  this problem does not occur in the first option
  since the modulus $|\det J(x,\tau)|$ is included in the 
integration measure $|d^Nz|$
in \eqref{eq:thible_integral_appendix3}
although the phase factor $e^{i\theta}= d^N z / |d^N z|$
should be taken into account by reweighting.\label{footnote:modulus}}
when the modulus $|\det J(x,\tau)|$
fluctuates considerably during the simulation.
In that case, only a small number of
configurations with large $|\det J(x,\tau)|$ dominate the
ensemble average and hence the statistics cannot be increased efficiently.
It turns out that this problem does not occur in the simulations performed
in this work if we optimize the flow equation as we describe in
section \ref{sec:optimized-flow}.
In all the simulations, we have chosen $\tau_{\rm min}=0.2$,
which is small enough to solve the multi-modality problem,
and $\tau_{\rm max}=4$,
which is large enough to obtain typical trajectories close to
the relevant saddle point.
Note also that the sign problem is solved already at $\tau \sim 2$.

Once we generate the configurations $(x , \tau)$,
we can calculate the expectation value $\langle \mathcal{O} \rangle$
by taking the ensemble average of $\mathcal{O}(z(x,\tau))$
with the reweighting factor 
\begin{align}
R(x,\tau) = \det J(x,\tau) \, e^{- i \, {\rm Im}S(z(x,\tau))}
\label{eq:reweithting-factor}
\end{align}
using the configurations
$( x , \tau)$
obtained for an appropriate range of $\tau$ \cite{Fukuma:2021aoo}.

In either option of the HMC algorithm, the most time-consuming
part is the calculation of the Jacobian $\det J(x,\tau)$,
which is needed only in the reweighting procedure.
In order to calculate the Jacobi matrix $J(x,\tau)$,
one has to solve the flow equation
\begin{equation}
  \frac{\partial}{\partial \sigma}J_{ij}(\sigma)
  =\overline{H_{ik}(z(\sigma))\, J_{kj}(\sigma)}
    \label{eq:Jacobian_floweq}
\end{equation}
with the initial condition $J(0)=\textbf{1}_N$,
where we have defined the Hessian
\begin{align}
  H_{ij}(z)
  &= \frac{\partial^2 S(z)}{\partial z_i\partial z_j} \ .
  \label{eq:def-Hessian}
\end{align}

\subsection{Optimizing the flow equation}
\label{sec:optimized-flow}

In this section we discuss a problem that occurs when we use
the original flow equation 
\eqref{anti-hol-grad-flow-general}
for a system with many variables such as 
the one studied below.\footnote{See
  Ref.~\cite{Feldbrugge:2022idb} for discussions on the
  gradient flow and its modification from a different point of view.}
We solve this problem by optimizing the flow equation, which
actually has large freedom of choice
if we are just to satisfy the property \eqref{prop-anti-hol-flow}.
Here we explain the basic idea and
defer a detailed discussion to the forth-coming paper \cite{precondition2022}.

The problem with the original flow
\eqref{anti-hol-grad-flow-general}
can be readily seen by considering how its solution $z(x,\sigma)$ changes
when the initial value $z(x,0) = x \in \bbR^N$ changes infinitesimally.
Note that the displacement 
$\zeta_i(\sigma)\equiv z_i(x+\delta x,\sigma)-z_i(x,\sigma)$
for an infinitesimal $\delta x$
can be obtained as
\begin{align}
\zeta_i(\sigma) = J_{ij}(\sigma) \, \delta x_j \ ,
\end{align}
where $J_{ij}(\sigma)$ is the Jacobi matrix at the flow time $\sigma$,
which satisfies the flow equation \eqref{eq:Jacobian_floweq}.
Thus we find that the displacement satisfies the flow equation
\begin{equation}
  \frac{d\zeta_i(\sigma)}{d\sigma}
    =\overline{H_{ij}(z(\sigma)) \, \zeta_j(\sigma)}
\label{eq:zeta-time-evolve}
\end{equation}
with the boundary condition $\zeta_i(0)=\delta x_i$,
where $H_{ij}(z)$ is the Hessian defined by \eqref{eq:def-Hessian}.

Let us consider the singular value decomposition (SVD)
of the Hessian $H_{ij}(z(\sigma))$ given as\footnote{This is known
  as the Takagi decomposition,
  which is the SVD for a complex symmetric matrix.}
\begin{align}
  H(z(\sigma)) = U^\top (\sigma) \, \Lambda(\sigma) \, U(\sigma) \ ,
\label{eq:H-SVD}
\end{align}
where $U(\sigma)$ is a unitary matrix and
$\Lambda={\rm diag}(\lambda_1 , \cdots , \lambda_N)$
is a diagonal matrix with $\lambda_1 \ge \cdots \ge  \lambda_N \ge 0$.
Plugging this in \eqref{eq:Jacobian_floweq},
we obtain
\begin{align}
  \frac{d J (\sigma)}{d\sigma}
  = U(\sigma)^\dag \, \Lambda(\sigma) \,
  \overline{U(\sigma) \, J(\sigma)} \ ,
\label{eq:Jacobian-time-evolve2}
\end{align}
and similarly for the displacement
\begin{align}
  \frac{d\zeta(\sigma)}{d\sigma}
  = U(\sigma)^\dag \, \Lambda(\sigma) \,
  \overline{U(\sigma) \, \zeta(\sigma)} \ .
\label{eq:zeta-time-evolve2}
\end{align}
  Roughly speaking, the magnitude of the
  displacement $\zeta(\sigma)$  grows exponentially
  with $\sigma$, and
  the growth rate
  is given by
  a weighted average of the singular values with a weight
  depending on $\delta x$.
If the singular values have a hierarchy $\lambda_1 \gg \lambda_{N}$,
some modes grow much faster than the others.
This causes a serious technical problem in solving the flow equation 
since it may easily diverge.
%

  In order to solve this problem, we pay attention to the freedom in
  defining the flow equation. As we discussed in section \ref{section:GTM},
  the important property of
  the flow equation \eqref{anti-hol-grad-flow-general}
  is \eqref{prop-anti-hol-flow}.
  Let us therefore consider a generalized flow equation
\begin{align}
    \frac{d z_i(\sigma)}{d\sigma}
    =\mathcal{A}_{ij}(z(\sigma),\overline{z(\sigma)})
      \overline{\frac{\partial S(z(\sigma))}{\partial z_j}} \ .
\label{anti-hol-grad-flow-general-opt}
\end{align}
Then the equation \eqref{prop-anti-hol-flow} becomes
\begin{equation}
    \frac{d S(z(\sigma))}{d\sigma}
=\sum_i \frac{\partial S(z(\sigma))}{\partial z_i}
    \frac{d z_i(\sigma)}{d \sigma} 
    =\sum_{ij} \frac{\partial S(z(\sigma))}{\partial z_i}
    \mathcal{A}_{ij}(z(\sigma),\overline{z(\sigma)})
\overline{\frac{\partial S(z(\sigma))}{\partial z_j}} \ .
\label{prop-anti-hol-flow-general}
\end{equation}
For this to be positive semi-definite,
the kernel $\mathcal{A}_{ij}(z,\bar{z})$ has only to be Hermitian positive,
and it does not have to be holomorphic.

The flow of the Jacobi matrix
becomes
\begin{equation}
  \frac{\partial}{\partial \sigma}J_{ij}(\sigma)
  =\mathcal{A}_{ik} \overline{H_{kl}(z(\sigma))\, J_{lj}(\sigma)}
  + \left(
  \frac{\partial \mathcal{A}_{il}}{\partial z_k} J_{kj}(\sigma)
  + \frac{\partial \mathcal{A}_{il}}{\partial \bar{z}_k} \overline{J_{kj}(\sigma)} \right)
    \overline{ \frac{\partial S(z(\sigma))}{\partial z_l}} 
 \ .
      \label{eq:Jacobian_floweq-preconditioned}
\end{equation}
From \eqref{eq:Jacobian_floweq-preconditioned},
we obtain the flow of the displacement as
\begin{equation}
  \frac{\partial}{\partial \sigma} \zeta_i(\sigma)
  =\mathcal{A}_{ik} \overline{H_{kl}(z(\sigma))\, \zeta_l(\sigma)}
  + \left(
  \frac{\partial \mathcal{A}_{il}}{\partial z_k} \zeta_k(\sigma)
  + \frac{\partial \mathcal{A}_{il}}{\partial \bar{z}_k} \overline{\zeta_k(\sigma)} \right)
    \overline{ \frac{\partial S(z(\sigma))}{\partial z_l}} 
 \ .
      \label{eq:zeta_floweq-preconditioned}
\end{equation}
Let us here assume that the first term is
dominant\footnote{This assumption is valid when $z(\sigma)$ is
  close to a saddle point, for instance. Otherwise, it should be 
simply regarded as a working hypothesis.}
in \eqref{eq:zeta_floweq-preconditioned}.
Then plugging \eqref{eq:H-SVD}
in \eqref{eq:zeta_floweq-preconditioned},
we obtain
\begin{align}
  \frac{d\zeta(\sigma)}{d\sigma}
  \sim \mathcal{A} \, U(\sigma)^\dag \, \Lambda(\sigma) \,
  \overline{U(\sigma) \, \zeta(\sigma)} \ .
\label{eq:zeta-time-evolve2-opt}
\end{align}
Therefore, by choosing 
\begin{align}
  \mathcal{A} = U(\sigma)^\dag \, \Lambda^{-1}(\sigma)\, U(\sigma) \ ,
  \label{eq:A-optimal}
  \end{align}
we
obtain
\begin{align}
  \frac{d\zeta(\sigma)}{d\sigma}
  \sim U(\sigma)^\dag \, \overline{U(\sigma) \, \zeta(\sigma)} \ ,
\label{eq:zeta-time-evolve3}
\end{align}
in which
the problematic hierarchy of
singular values $\lambda_i$ in \eqref{eq:zeta-time-evolve2}
is completely eliminated.
From this point of view, \eqref{eq:A-optimal} seems to be the optimal choice
for the ``preconditioner'' $\mathcal{A}$ 
in the
generalized flow equation \eqref{anti-hol-grad-flow-general-opt}.
Note also that, under a similar assumption, the flow of the Jacobi matrix
changes from \eqref{eq:Jacobian-time-evolve2} to
\begin{equation}
  \frac{\partial}{\partial \sigma}J(\sigma)
  \sim
   U(\sigma)^\dag \, \overline{U(\sigma) \, J(\sigma)} \ .
      \label{eq:Jacobian_floweq-preconditioned-2}
\end{equation}
Therefore, the use of the optimal flow equation
reduces
the overlap problem that occurs
due to the large fluctuation of $|\det J|$ in our algorithm.

In order to implement this idea in the simulation, let us first note 
that \eqref{eq:A-optimal} can be written as
\begin{align}
  \mathcal{A}(z(\sigma),\overline{z(\sigma)}) 
  &=
  \Big\{ H^\dag(z(\sigma)) H(z(\sigma)) \Big\}^{-1/2}
  = \Big\{ \overline{H(z(\sigma))} H(z(\sigma)) \Big\}^{-1/2} \ .
  \label{eq:A-optimal-2}
  \end{align}
Here we use the rational approximation 
\begin{align}
x^{-1/2} 
&\approx
a_0  + \sum_{q=1}^Q \frac{a_q}{ x + b_q } \ ,
\label{eq:rational_approximation}
\end{align}
which can be made accurate for a wide range of $x$
with the real positive parameters $a_q$ and $b_q$
generated by the Remez algorithm.
Thus we obtain
\begin{align}
    \mathcal{A}(z,\bar z)
    &\approx a_0 \, \textbf{1}_N
    + \sum_{q=1}^Q a_q \, \Big\{ \overline{H(z)} H(z)+b_q\, \textbf{1}_N \Big\}^{-1} \ ,
    \label{eq:rational_approximation-2}
\end{align}
which is much easier to handle on a computer
than \eqref{eq:A-optimal}.
In particular, the matrix inverse $(\bar HH+b_q\textbf{1}_N)^{-1}$ does not have to
be calculated explicitly since it only appears in the algorithm 
as a matrix that acts on a particular vector, which allows us to
use an iterative method for solving a linear equation
such as the conjugate gradient (CG) method.
The factor of $Q$ in the computational cost
can be avoided by the use of a multi-mass CG solver \cite{Jegerlehner:1996pm}.
These techniques are well known 
in the so-called Rational HMC
  algorithm \cite{Kennedy:1998cu,Clark:2006wq},
which is widely used in
  QCD with dynamical strange quarks \cite{Clark:2004cp} and
  supersymmetric theories such as the BFSS and IKKT matrix 
models (See Refs.~\cite{Catterall:2007fp,Anagnostopoulos:2007fw,Kim:2011cr}, 
for example.).


\section{Monte Carlo results obtained by the GTM}
\label{section:complex_traj_from_montecarlo}

In this section we present our results
obtained by Monte Carlo calculations based on the GTM.
The path integral
for the time-evolved wave function
can be represented as\footnote{Here we
  omit the normalization factor
for the wave function, which will not be important throughout this article.}
\begin{equation}
  \Psi(x_{\rm f} ; T) = \int dx
\,  
e^ {-S_{\rm eff}(x)} \ ,
    \label{eq:time_evolution_path_integral-discrete}
\end{equation}
where $dx = \prod_{n=0}^{N-1} dx_{n}$.
The effective action 
$S_{\rm eff}(x)$ is
a function of $x=(x_0 , \cdots , x_{N-1})$ given by\footnote{Note
that the log term in \eqref{eq:eff-action} has a branch cut.
This does not cause any problem below, however,
since in actual calculations we only need either
$\partial S_{\rm eff}(x)/\partial x$
or $\exp{(-S_{\rm eff}(x))}$.\label{footnote:log-action}}
\begin{equation}
    S_{\rm eff}(x)=
-\frac{i \epsilon}{\hbar} \sum_{n=0}^{N-1}\left\{
\frac{1}{2} \, m \left( \frac{x_{n+1}-x_n}{\epsilon}\right)^2
-\frac{V(x_{n+1})+V(x_n)}{2}\right\}
-\log\Psi(x_0)  \ ,
\label{eq:eff-action}
\end{equation}
where $x_N = x_{\rm f}$
and the initial wave function
is chosen as
\begin{equation}
  \Psi (x) = 
  \frac{1}{(2\pi)^{1/4}\sigma^{1/2}}
  \exp\left\{-\frac{1}{4\sigma^2}(x-b)^2 + \frac{ipx}{\hbar} 
  \right\} \ .
    \label{eq:gaussian_ansatz}
\end{equation}
%
In all the simulations in this article,
we set $m=1$, $\hbar=1$,
and the total time to $T=2$,
which is
divided into $N=20$ intervals.

Here we consider the double-well potential
\eqref{eq:double-well_potential} with $a=1$.
The height of the potential at the local maximum $x=0$ is $V_0=\lambda$.
We use $b=-1$ and $\sigma=0.3$
for the initial wave function \eqref{eq:gaussian_ansatz}
so that it is well localized around $x=-1$, which is one of the potential minima.
We choose the end point
$x_{\rm f}=1$ to be the other potential minimum.

In order to choose an appropriate value for $\lambda$ in the potential
to probe quantum tunneling,
we consider the probability 
\begin{equation}
    P =\sum_{E \geq V_0} |\langle E|\Psi\rangle|^2 
    \label{eq:classical_tunneling_probability}
\end{equation}
of the initial quantum state having energy larger than the
potential barrier $V_0$, where $| E \rangle$ represents the
normalized energy eigenstate with the energy $E$.
If we choose the momentum $p=0$ in
the initial wave function \eqref{eq:gaussian_ansatz},
we obtain $P \sim 0.11$ for $\lambda=2.5$.
We therefore use $\lambda=2.5$ in our calculation.

\begin{figure}
    \centering
    \includegraphics[scale=0.8]{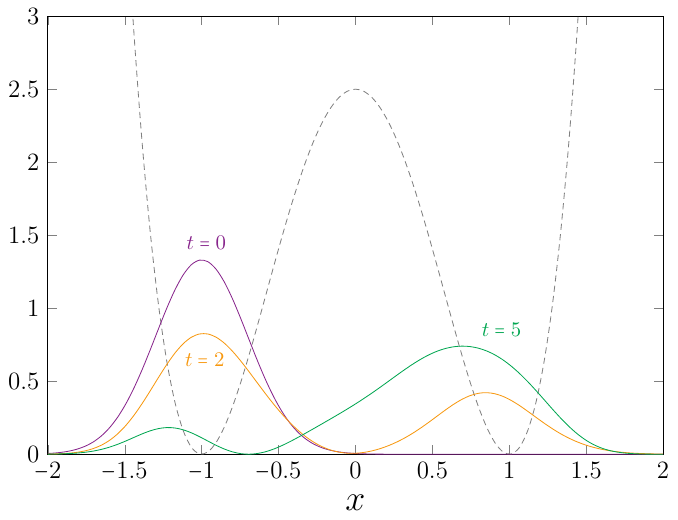}
    \caption{The distribution $|\Psi(x;t)|^2$
      at time $t=0$ (purple line), $t=2$ (yellow line) and $t=5$ (green line)
      are shown for the initial wave function \eqref{eq:gaussian_ansatz}
      with $\sigma=0.3$, $b=-1$, $p=0$
      in the double-well potential \eqref{eq:double-well_potential}
      with $\lambda=2.5$, $a=1$ (gray, dashed line).}
    \label{fig:example_Pt}
\end{figure}

\begin{figure}
    \centering
    \includegraphics[scale=0.75]{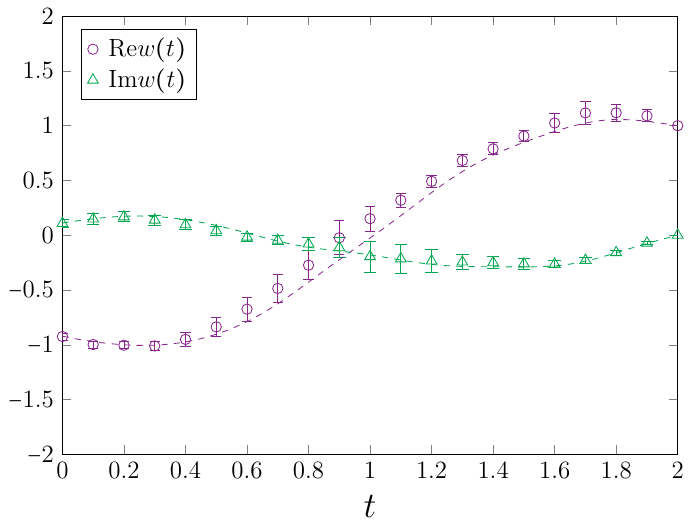}
    \includegraphics[scale=0.75]{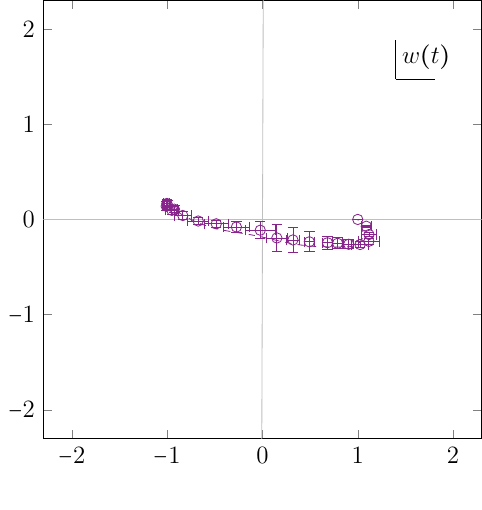}
    \includegraphics[scale=0.75]{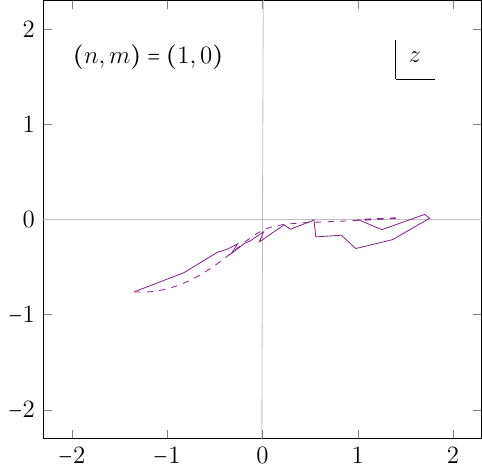}
    \includegraphics[scale=0.75]{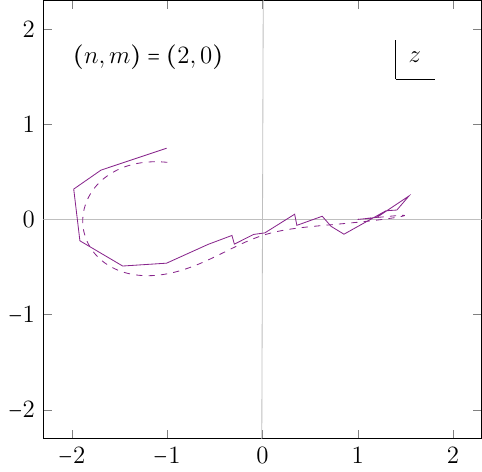}
    \caption{The results obtained
      for the initial wave function \eqref{eq:gaussian_ansatz}
      with $\sigma=0.3$, $b=-1$, $p=0$
      and
      $x_{\rm f}=1$ 
      in the double-well potential \eqref{eq:double-well_potential}
      with $\lambda=2.5$, $a=1$.
      (Top) The weak value of the coordinate is plotted against time $t$
      in the Left panel, while the trajectory of the weak value is plotted 
      in the complex plane in the Right panel.
      The dashed lines represent the result obtained
      from \eqref{def-weak-value} by solving the Schr\"odinger equation.
(Bottom) Two typical trajectories obtained from the numerical simulation
      with the same parameters as in the Top panels.
   The dashed lines represent the closest classical solutions 
   obtained by choosing the mode $(n,m)$ and the initial point $x_{\rm i}$
   with the final point $x_{\rm f}=1$ fixed.
}
    \label{fig:ctrl_WM}
\end{figure}


\begin{figure}
    \centering
    \includegraphics[scale=0.75]{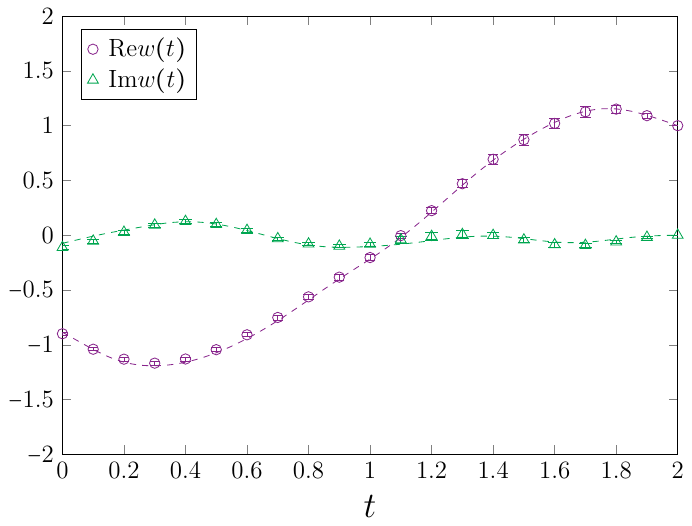}
    \includegraphics[scale=0.75]{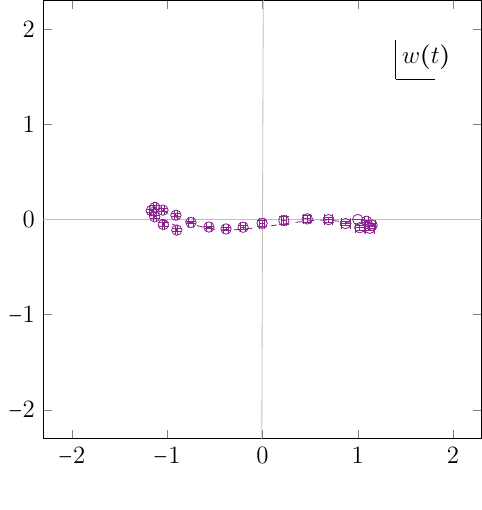}
    \includegraphics[scale=0.75]{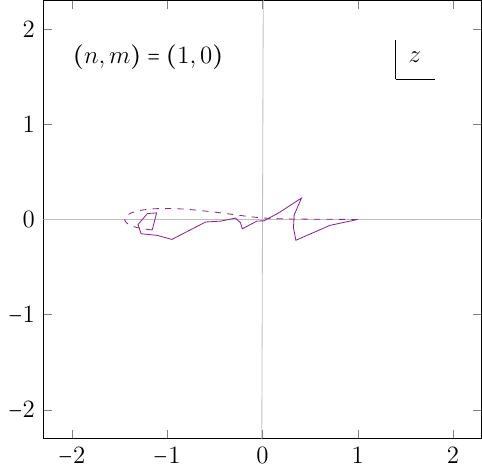}
    \includegraphics[scale=0.75]{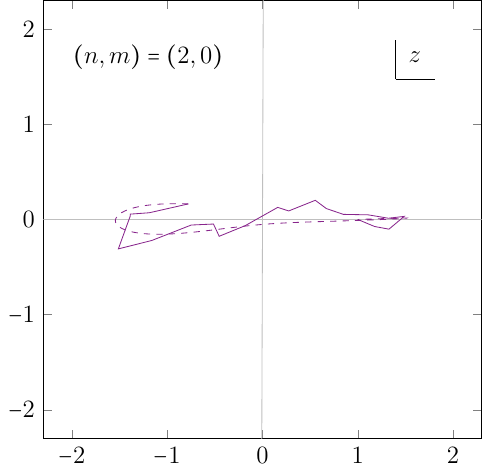}
    \caption{The results obtained
      for the initial wave function \eqref{eq:gaussian_ansatz}
      with the momentum $p=-2$. All the other parameters are the same
      as in Fig.~\ref{fig:ctrl_WM}.
      (Top) The weak value of the coordinate is plotted against time $t$
      in the Left panel, while the trajectory of the weak value is plotted 
      in the complex plane in the Right panel.
      The dashed lines represent the result obtained
      from \eqref{def-weak-value} by solving the Schr\"odinger equation.
      (Bottom) Two typical trajectories obtained from the numerical simulation
         with the same parameters as in the Top panels.
   The dashed lines represent the closest classical solutions 
   obtained by choosing the mode $(n,m)$ and the initial point $x_{\rm i}$
   with the final point $x_{\rm f}=1$ fixed.}
    \label{fig:k2_WM}
\end{figure}

Note that a typical tunneling time can be evaluated by
\begin{equation}
    t_0 \sim\frac{\pi\hbar}{\Delta E} \ ,
\end{equation}
where $\Delta E$ is the energy difference between
the ground state and the first excited state.
For $\lambda=2.5$, we find $t_0 \sim 5$.
In Fig.~\ref{fig:example_Pt} we plot the wave functions
at $t=0$, $t=2$ and $t=5$
obtained for this setup 
by solving the Schr\"odinger equation 
with Hamiltonian diagonalization.
The result for $t=2$ shows that the significant portion of the distribution
has moved to the other potential minimum $x=1$,
which implies that quantum tunneling has indeed occurred.

The expectation value of the coordinate $x(t)$ at time $t$
in the path integral \eqref{eq:time_evolution_path_integral-discrete}
gives the weak value defined as \cite{Aharonov:1988xu}
\begin{alignat}{3}
w(t) &= \frac{\langle x_{\rm f} | \, \hat{U}(T-t) \, \hat{x} \, 
\hat{U}(t) \, | \Psi \rangle}
{\langle x_{\rm f}  | \, \hat{U}(T) \, |\Psi \rangle} \ , 
\label{def-weak-value}
\end{alignat}
where 
$\hat{U}(t) = \exp(-i \, t\hat{H}/\hbar)$ is 
the time-evolution unitary operator with the Hamiltonian $\hat{H}$.
The quantum state $|\Psi \rangle$ corresponds to the
initial wave function and $| x_{\rm f} \rangle$ represents
the eigenstate of the coordinate operator\footnote{In general, the weak value
can be defined for an arbitrary post-selected final state $|\Phi \rangle$ instead of $| x_{\rm f} \rangle$.}.
Note that 
the weak value 
is complex in general
unlike the expectation value, which is real
for a Hermitian operator such as $\hat{x}$.
It is not only a mathematically well-defined quantity
but also a physical quantity that can be measured by experiments
using the so-called ``weak measurement'' \cite{Aharonov:1988xu}.
Below we will see that the effects of the complex saddle points can be 
probed by the ``weak value'' of the coordinate operator $\hat{x}$
as pointed out in Refs.~\cite{TANAKA2002307,Turok:2013dfa}.

In Fig.~\ref{fig:ctrl_WM} (Top), we show our results for
the weak value $w(t)$ of the coordinate at time $t$ 
defined by \eqref{def-weak-value}
for the initial wave function \eqref{eq:gaussian_ansatz}
with $\sigma=0.3$, $b=-1$, $p=0$
and $x_{\rm f}=1$
     in the double-well potential \eqref{eq:double-well_potential}
      with $\lambda=2.5$, $a=1$.
      The dashed lines represent the results obtained 
      directly from \eqref{def-weak-value}
      by solving the Schr\"odinger equation with Hamiltonian diagonalization.
      The agreement between our data and the direct results 
confirms the validity of our calculation.
We find that the weak value $w(t)$ is indeed complex 
except for the end point, which is fixed to $w(T)=x_{\rm f}=1$.
Note that $w(0)$ is also complex although it is
close to $x=-1$, which is the center of 
the Gaussian wave function \eqref{eq:gaussian_ansatz}.

In the Bottom panels of Fig.~\ref{fig:ctrl_WM},
 we show two typical trajectories
obtained from the simulation with the same parameters as 
in the Top panels.
These trajectories are obtained for a relatively long
flow time $\tau=4$ in the 
GTM (See section \ref{section:integrating-flow-time}.)
and therefore they are expected to
be close to some relevant saddle points.
Indeed we are able to find a classical solution discussed in
section \ref{sec:exact-solutions}, which is close
to each of these trajectories
by choosing the mode $(n,m)$ and the initial point $x_{\rm i}$
with the final point $x_{\rm f}=1$ fixed.
We find that the typical trajectories have a larger imaginary part
on the left and a smaller imaginary part on the right,
which suggests that quantum tunneling occurs first and then 
some classical motion follows.
Note that this is different from the behaviors of the weak value $w(t)$ 
shown in the Top-Right panel.
This is due to the fact that the weak value $w(t)$ is a weighted average
of $x(t)$ obtained from the simulation, where the 
weight \eqref{eq:reweithting-factor} is complex
in general since it consists of the phase factor $e^{-i{\rm Im}S_{\rm eff}}$
and the Jacobian for the change of 
variables.


Next we introduce nonzero momentum
$p=-2$ in the initial wave function \eqref{eq:gaussian_ansatz}.
In Fig.~\ref{fig:k2_WM} we show our results
with all the other parameters the same as in Fig.~\ref{fig:ctrl_WM}.
Since the initial kinetic energy is $p^2/2 = 2$, which is close to
the potential barrier $\lambda = 2.5$, 
a classical motion over the potential barrier is
possible if the initial point $x(0)$ is slightly shifted
from the potential minimum.
Indeed we find that the weak value
and the typical trajectories
become close to real.

\section{Summary and discussions}
\label{section:summary}

We have investigated 
quantum tunneling in the real-time path integral \cite{Nishimura:2023dky},
which has important applications in QFT, quantum cosmology and so on.
Unlike the previous works, we performed explicit Monte Carlo calculations
based on the GTM.
In particular, we were able to identify
the complex trajectories
that are relevant from the viewpoint of the Picard-Lefschetz theory.
When we introduce momentum in the initial wave function,
we found that the trajectories come closer to real,
which clearly indicates the transition to classical dynamics.

In actual calculations,
we integrate
the flow time
within an appropriate range
to overcome the sign problem and the multi-modality problem \cite{Fukuma:2019uot}.
We use the HMC algorithm
on the real axis instead of using it on the deformed contour.
This is made feasible by calculating the force using
backpropagation \cite{Fujisawa:2021hxh}.
Optimizing the flow equation \cite{precondition2022} is also important.
In particular,
we do not observe the overlap problem due to reweighting $|{\rm det} J|$
in our calculations.
We hope that our approach is useful
in studying the real-time dynamics of various quantum systems.


\subsection*{Acknowledgements}

We would like to thank Katsuta Sakai and Atis Yosprakob for
collaborations \cite{Nishimura:2023dky,precondition2022}
that produced the results reported in this article.
We are also grateful to Yuhma Asano and Masafumi Fukuma
for valuable discussions.
The computations were carried out on
the PC clusters in KEK Computing Research Center
and KEK Theory Center.

\bibliographystyle{JHEP}
\bibliography{ref}



\end{document}